\documentclass[12pt]{iopart}

\usepackage{graphicx}
\usepackage{subfigure}
\usepackage{verbatim}
\usepackage{dcolumn}% Align table columns on decimal point
\usepackage{bm}% bold math
\usepackage{epsf}
\usepackage{color}
\usepackage[colorlinks=true,citecolor=blue,linkcolor=blue,urlcolor=blue]{hyperref}
\usepackage{hhline}
\usepackage{amssymb}
\usepackage{cite}

  \expandafter\let\csname equation*\endcsname\relax
  \expandafter\let\csname endequation*\endcsname\relax
\usepackage{amsmath}

\newcommand{\pd}{\partial}

\newcommand{\td}{\mathrm{d}}

\newcommand{\ex}[1]{\exp{\left(#1\right)}}
\newcommand{\id}{\mathbb{I}}

\newcommand{\si}[1]{\sin{\left(#1\right)}}

\newcommand{\bla}{bla\\bla\\bla\\bla\\bla}

\newcommand{\mb}[1]{\mbox{\boldmath$#1$}}

\newcommand{\mbb}[1]{\mathbb{#1}}

\newcommand{\mrm}[1]{\mathrm{#1}}

\begin{document}

\title{Shortcuts to adiabaticity: Suppression of pair production in driven Dirac dynamics}

\author{Sebastian Deffner}
\address{Theoretical Division and Center for Nonlinear Studies, Los Alamos National Laboratory, Los Alamos, NM 87545, USA}
\ead{sebastian.deffner@gmail.com}

\date{\today}

\begin{abstract}
Achieving effectively adiabatic dynamics in finite time is a ubiquitous goal in virtually all areas of modern physics. So-called shortcuts to adiabaticity refer to a set of methods and techniques that allow to produce in a short time the same final state that would result from an adiabatic, infinitely slow process. In the present work we generalize one of these methods -- the fast-forward technique -- to driven Dirac dynamics. As a main result we find that shortcuts to adiababticity for the $(1+1)$-dimensional Dirac equation are facilitated by a combination of both, scalar and pseudoscalar potentials. Our findings are illustrated for two analytically solvable examples, namely charged particles driven in spatially homogeneous and linear vector fields.
\end{abstract}


%\maketitle

\section{Introduction}

Although initially not even fully appreciated by its discoverer the ``Dirac equation'' \cite{Dirac1928} has turned out to be one of the most important and versatile results of theoretical physics. In particle physics, the Dirac equation is a relativistic wave equation, which describes massive spin-$1/2$ particles, such as electrons and quarks \cite{Thaller1956,Peskin1995}. Among its achievements are the explanation of spin as a consequence of special relativity and the discovery of the existence of antimatter. Even almost a century after its inception the analysis of Dirac dynamics and the enhancement of pair production -- the creation of antimatter-- still attracts vivid research activity \cite{Pickl2008,Fillion2012,Fillion2013,Fillion2013a}.

However, study and applications of the Dirac equation are no longer only restricted to problems in particles physics. For instance, Dirac dynamics have been recently analyzed in the context of the quantum speed limit \cite{Villamizar2015}, in optomechanics \cite{Schmidt2015}, and in quantum thermodynamics \cite{Deffner2015}. More importantly, however, in a wide rage of materials, such as $d$-wave superconducters and graphene, low-energy excitations behave as massless Dirac particles rather than fermionic Schr\"odinger particles. So-called ``Dirac materials''  share many universal properties and are subject to intense experimental and theoretical research \cite{Wehling2014}. Similarly to electron-positron pair production in particles physics it has been studied how the pair-creation can be enhanced in Dirac materials \cite{Fillion2015}. 

All mechanisms to enhance pair-production, in particle physics as well as in condensed matter physics, have to overcome the limitations rooted in the quantum adiabatic theorem for Dirac dynamics \cite{Nenciu1980,Nenciu1980a,Nenciu1981,Faisal2011}. In the present analysis we are interested in the reverse problem. Imagine, e.g., a situation in which a Dirac material is subjected to an externally controlled electric field. For finite-time variation of such a field electron-hole pairs are created.  For instance, in superconductors these electron-hole pairs can be described as quasiparticles, which are normal conducting and experience Joule heating. Such nonequilibrium excitations are often undesirable, but they are an inevitable consequence of finite-time driving. In this case one is interested in particular finite-time processes that mimic adiabatic evolution -- shortcuts to adiabaticity. In recent years a great deal of theoretical and experimental research has been dedicated to the  design of shortcuts to adiabaticity for Schr\"odinger dynamics \cite{Torrontegui2013}. To this end a variety of techniques has been developed: the use of dynamical invariants \cite{Chen2010}, the inversion of scaling laws \cite{Campo2012}, the fast-forward technique \cite{Masuda2010,Masuda2011,Torrontegui2012,Torrontegui2012a,Masuda2014a,Kiely2015}, transitionless quantum driving \cite{Demirplak2003,Demirplak2005,Berry2009,Deffner2014}, optimal protocols from optimal control theory \cite{Chen2011a,Stefanatos2013,Campbell2014}, optimal driving from properties of the quantum work statistics  \cite{Xiao2014}, ``environment'' assisted methods \cite{Masuda2014}, using the properties of Lie algebras \cite{Torrontegui2014}, and approximate methods such as linear response theory \cite{Acconcia2015} and fast quasistatic dynamics \cite{Garaot2015}. 

Among these methods the fast-forward technique stands out. Within this approach one determines an auxiliary potential such that the time-dependent wave function becomes identical to a ``target state'' at the end of the finite-time process \cite{Torrontegui2012a}. For intermediate times, however, the time-dependent state is allowed to stray arbitrarily far from the adiabatic manifold. This is in stark contrast to, for instance, counterdiabatic driving for which the time-dependent solution remains on the adiabatic manifold of the ``unperturbed" Hamiltonian at all times \cite{Berry2009}. However, the counterdiabatic fields tends to be highly complicated and highly non-local \cite{Deffner2014,Campbell2014}, whereas the auxiliary fast-forward potential is more tractable \cite{Masuda2014a}. Therefore, generalizing the fast-forward technique to facilitate shortcuts to adiabatictity for Dirac dynamics appears to be more promising than, for instance, counterdiabatic driving. Generally, one would expect optimal control techniques to require even less resources that the fast-forward technique. However, solving optimal control problems can be become rather mathematically involved \cite{Deffner2014d}. Thus, the present work focuses on a mathematically rather ``simple'', yet effective method -- the fast forward technique.

In the following we will briefly review the main properties of Dirac dynamics in Sec.~\ref{sec:dir}, before we will develop the fast-forward technique for Dirac particles in time-dependent fields in Sec.~\ref{sec:FF}. We will see that shortcuts to adiabaticity for Dirac dynamics are facilitated by pseudoscalar potentials, which will be illustrated for two analytically solvable examples in Sec.~\ref{sec:ex}.

\section{Relativistic quantum mechanics: The Dirac equation \label{sec:dir}}

We start by briefly reviewing the main properties of the Dirac dynamics. In its original formulation  for free particles the Dirac equation reads \cite{Dirac1928,Thaller1956}
\begin{equation}
\label{eq01}
i\hbar\, \dot{\Psi}(\mb{x},t)=\left(-i\hbar c\, \mb{\alpha}\cdot\nabla+\alpha_0\,mc^2\right)\,\Psi(\mb{x},t)\,.
\end{equation}
Here, $\Psi(\mb{x},t)$ is the wave function of an electron with rest mass $m$ at position $\mb{x}=(x_1,x_2,x_3)$, and $c$ is the speed of light. In covariant form the matrices $\mb{\alpha}=(\alpha_1,\alpha_2,\alpha_3)$ and $\alpha_0$ can be expressed as \cite{Peskin1995,Thaller1956},
\begin{equation}
\label{eq02}
\alpha^0=\gamma^0\quad\mathrm{and}\quad\gamma^0\, \alpha^k=\gamma^k\,.
\end{equation}
The $\gamma$-matrices are commonly written in terms of $2\times 2$ sub-matrices with the Pauli-matrices $\sigma_x, \sigma_y, \sigma_z$ and the identity $\id_2$ as,
\begin{eqnarray}
\label{eq03}
\fl
\gamma^0=\begin{pmatrix}\id_2&0\\0&-\id_2 \end{pmatrix}\quad&\gamma^1=\begin{pmatrix}0&\sigma_x\\-\sigma_x&0 \end{pmatrix}\quad
\gamma^2=\begin{pmatrix}0&\sigma_y\\-\sigma_y&0 \end{pmatrix}\quad&\gamma^3=\begin{pmatrix}0&\sigma_z\\-\sigma_z&0 \end{pmatrix}\,.
\end{eqnarray}
The solution of Eq.~\eqref{eq01}, the Dirac wave function $\Psi(\mb{p},t)$, is a bispinor, which can be interpreted as a superposition of a spin-up electron, a spin-down electron, a spin-up positron, and a spin-down positron \cite{Thaller1956,Peskin1995}. \footnote{The momentum of Dirac particles is confined by the light cone, whereas Schr\"odinger particles can travel with arbitrary velocities. This has interesting consequences for the quantum work statistics \cite{Deffner2015}.}

We will now be interested in situations in which Dirac particles are driven by a time-dependent vector field $\mb{A}(\mb{x},t)$. Note that considering only situations with time-dependent vector field, but no scalar potential is not a loss of generality, since electric and magnetic fields, $\mb{E}(\mb{x},t)$ and $\mb{B}(\mb{x},t) $, are gauge invariant. In particular, the physical fields are given by
\begin{equation}
\label{eq04}
\mb{E}(\mb{x},t)=-\nabla\, V(\mb{x},t)-\pd_t \mb{A}(\mb{x},t)\quad \mathrm{and}\quad \mb{B}(\mb{x},t)=\nabla \times \mb{A}(\mb{x},t)\,.
\end{equation}
Furthermore, for the sake of simplicity we will restrict ourselves in the following analysis to a one-dimensional system in $x$-direction. In this case the 4-component Dirac spinor can be separated into two identical 2-component bispinors, and we choose a representation in which the $(1+1)$-dimensional Dirac-equation reads \cite{deCastro2003,Solomon2010},
\begin{equation}
\label{eq05}
i\hbar\, \dot{\Psi}(x,t)=\left[\left(-i\hbar c\, \pd_x+A(x,t)\right)\,\sigma_x+ m c^2\, \sigma_z\right]\,\Psi(x,t)\,.
\end{equation}
For time-dependent, but spatially homogeneous vector fields this situation has been recently solved analytically for oscillating $A(t)$  in Ref.~\cite{Fillion2012} and for linear and exponential parametrization in Ref.~\cite{Deffner2015}. In addition, Eq.~\eref{eq05} describes particle-antiparticle production in counterpropagating laser light, which has been proposed to be observable in an experiment \cite{Fillion2012}. 

\section{Fast-forward technique \label{sec:FF}}

In the following we will generalize the fast-forward technique to facilitate shortcuts to adiabaticity for the time-dependent Dirac equation \eref{eq05}. We begin by reviewing notions and by establishing notations for the corresponding Schr\"odinger case.

\subsection{Schr\"odinger dynamics}

In the paradigm of the fast-forward technique \cite{Masuda2010,Masuda2011,Torrontegui2012a,Masuda2014a,Kiely2015} we are interested in finding a shortcut for the ``unperturbed'' time-dependent dynamics,
\begin{equation}
\label{eq06}
i \hbar\, \pd_t \psi_0(x,t)=\frac{1}{2 m}\left(-i \hbar\, \pd_x +\frac{1}{c}\, A(x,\alpha_t)\right)^2\,\psi_0(x,t)\,,
\end{equation}
where $\psi_0(x,t)$ denotes the ``unperturbed'', non-adiabatic solution and the vector field is controlled externally according to some protocol $\alpha_t$. Now, we consider an ansatz of a time-dependent wave function,
\begin{equation}
\label{eq07}
\psi(x,t)=\ex{-\frac{i}{\hbar}\int_0^t\td s\,\epsilon(\alpha_s)}\,\ex{i\,f(x,t)}\,\phi(x,\alpha_t)\,,
\end{equation}
where  $\epsilon(\alpha_t)$ is the instantaneous eigenenergy corresponding to an instantaneous eigenstate $\phi(x,\alpha_t)$,
\begin{equation}
\label{eq08}
\epsilon(\alpha_t)\, \phi(x,\alpha_t)=\frac{1}{2 m}\left(-i \hbar\, \pd_x +\frac{1}{c}\, A(x,\alpha_t)\right)^2\,\phi(x,\alpha_t)\,.
\end{equation}
The task is to determine the phase $f(x,t)$ and the auxiliary scalar potential $V(x,t)$ such that $\psi(x,t)$ \eref{eq07} is a solution of
\begin{equation}
\label{eq09}
i \hbar\, \pd_t \psi(x,t)=\frac{1}{2 m}\left(-i \hbar\, \pd_x +\frac{1}{c}\, A(x,\alpha_t)\right)^2\,\psi(x,t)+V(x,t)\,\psi(x,t)\,.
\end{equation}
A shortcut to adiabaticity during finite time $\tau$ is achieved if we additionally have $\psi(x,0)=\phi(x,\alpha_0)$ and $\psi(x,\tau)=\ex{-i/\hbar\,\int_0^\tau\td s\,\epsilon(\alpha_s)}\phi(x,\alpha_\tau)$, that is $f(x,0)=0$ and $f(x,\tau)=0$. Therefore, $\psi(x,\tau)$ becomes identical to the final state of adiabatically driving Eq.~\eref{eq06}, that means $\psi(x,\tau)$ is given by the instantaneous eigenstate times a space-independent phase.

For the following analysis it will prove convenient to express the instantaneous eigenstate in polar representation,
\begin{equation}
\label{eq10}
\phi(x,\alpha_t)=\beta(x,\alpha_t)\,\ex{i\, \gamma(x,\alpha_t)}
\end{equation}
with real amplitude $\beta(x,\alpha_t)$ and phase $\gamma(x,\alpha_t)$. Substituting the ansatz \eref{eq07} into the modified Schr\"odinger equation \eref{eq09} we obtain after a few lines
\begin{equation}
\label{eq11}
\begin{split}
V(x,t)&=-\hbar\, \pd_t f(x,t)-\hbar\,\pd_t\gamma(x,\alpha_t)+\frac{\hbar}{2 m c}\,A(x,\alpha_t)\,\pd_x f(x,t)\\
&+ \frac{\hbar^2}{2 m}\,\left[\pd_x f(x,t)\right]^2+\frac{\hbar^2}{m}\,\pd_x f(x,t)\,\pd_x\gamma(x,\alpha_t)\,.
\end{split}
\end{equation}
The time-dependent phase $f(x,t)$ is determined by the second-order differential equation,
\begin{equation}
\label{eq12}
\hbar\,\beta(x,\alpha_t)\,\pd_x^2f(x,t)+2 \hbar\,\pd_x\beta(x,\alpha_t)\,\pd_xf(x,t)-2 m \,\pd_t\beta(x,\alpha_t)=0\,.
\end{equation}
It is then easy to see that a shortcut to adiabaticity is obtained for all parametrizations with $\dot{\alpha}_0=0$ and $\dot{\alpha}_\tau=0$ \cite{Masuda2014a}, where the dot is a short notation for a derivative with respect to time 

It is interesting to note that the auxiliary potential $V(x,t)$ explicitly depends on the eigenstate $\phi(x,\alpha_t)$. This is in stark contrast to counterdiabatic driving, for which the counterdiabatic field constitutes a shortcut to adiabaticity for all eigenstates \cite{Deffner2014}.

\subsection{Dirac dynamics}

In complete analogy to the previous case of Schr\"odinger dynamics \eref{eq06} we are now interested in constructing a shortcut to adiabaticity for the time-dependent Dirac problem,
\begin{equation}
\label{eq13}
i\hbar\, \pd_t\,\Psi_0(x,t)=\left[\left(-i\hbar c\,\pd_x+A(x,\alpha_t)\right)\,\sigma_x+m c^2 \sigma_z\right]\,\Psi_0(x,t)\,,
\end{equation}
where the vector field, $A(x,\alpha_t)$, is again parametrized by a control parameter $\alpha_t$. As before \eqref{eq07}, we consider an ansatz
\begin{equation}
\label{eq14}
\Psi(x,t)=\ex{-\frac{i}{\hbar}\int_0^t\td s\,\epsilon(\alpha_s)}\,\ex{i\,f(x,t)}\,\Phi(x,\alpha_t)\,,
\end{equation}
where $\Phi(x,\alpha_t)=(\phi_1(x,\alpha_t), \phi_2(x,\alpha_t))$ is an eigenspinor with
\begin{equation}
\label{eq15}
\epsilon(\alpha_t)\,\Phi(x,\alpha_t)=\left[\left(-i\hbar c\,\pd_x+A(x,\alpha_t)\right)\,\sigma_x+m c^2\,\sigma_z\right]\Phi(x,\alpha_t)\,.
\end{equation}
In complete analogy to the Schr\"odinger case the task is now to determine phase, $f(x,t)$, and an auxiliary potential matrix, $\mbb{V}(x,t)$, such that $\Psi(x,t)$ \eref{eq14} is a solution of
\begin{equation}
\label{eq16}
i\hbar\, \pd_t\Psi(x,t)=\left[\left(-i\hbar c\,\pd_x+A(x,\alpha_t)\right)\,\sigma_x+m c^2 \sigma_z+\mbb{V}(x,t)\right]\,\Psi(x,t)\,.
\end{equation}
In the most general case the Lorentz invariant, hermitian potential matrix, $\mbb{V}(x,t)$, can be written as \cite{Thaller1956,Solomon2010},
\begin{equation}
\label{eq17}
\mbb{V}(x,t)=\id_2\,V_t(x,t)+ \sigma_x\, V_e(x,t)+\sigma_y\, V_p(x,t)+\sigma_z V_s(x,t)\,.
\end{equation}
Here, $V_t$ and $V_e$ are the time and space components of the 2-vector potential, $V_s$ is the scalar potential, and $V_p$ denotes the pseudoscalar potential \cite{Thaller1956,Castro2003,Haouat2007,Haouat2008,Solomon2010}. The existence of pseudoscalar potentials is an important consequence of Dirac dynamics \cite{Thaller1956}, and such potentials can be obtained by means of the ``inverse scatting method'' \cite{Toyama1999}. Physically, pseudoscalar potentials are induced by circularly polarized fields \cite{Finalli2009}.

As before, we work in a gauge such that $V_e(x,t)\equiv 0$, i.e., the space component of the vector potential is fully determined by the ``unperturbed'' problem \eref{eq13}. Substituting the ansatz \eref{eq15} into the modified evolution equation \eref{eq16} and rearranging terms we obtain from the first component,
\begin{equation}
\label{eq18}
\begin{split}
V_t(x,t)\, \phi_1(x,\alpha_t)=&-V_s(x,t)\,\phi_1(x,\alpha_t)+i V_p(x,t)\,\phi_2(x,\alpha_t)-\hbar\,\phi_1(x,\alpha_t)\, \pd_t f(x,t)\\
&\quad-c\hbar\,\phi_2(x,\alpha_t)\,\pd_x f(x,t)+i \hbar\, \pd_t \phi_1(x,\alpha_t)\,,
\end{split}
\end{equation}
and from the second component
\begin{equation}
\label{eq19}
\begin{split}
V_t(x,t)\, \phi_2(x,\alpha_t)=&\,V_s(x,t)\,\phi_2(x,\alpha_t)-i V_p(x,t)\,\phi_1(x,\alpha_t)-\hbar\,\phi_2(x,\alpha_t)\, \pd_t f(x,t)\\
&\quad -c\hbar\,\phi_1(x,\alpha_t)\,\pd_x f(x,t)+i \hbar\, \pd_t \phi_2(x,\alpha_t)\,.
\end{split}
\end{equation}
Multiplying Eq.~\eref{eq18} with $\phi_2(x,\alpha_t)$ and Eq.~\eref{eq19} with $\phi_1(x,\alpha_t)$, and subtracting the resulting equations we have,
\begin{equation}
\label{eq20}
\begin{split}
&c\hbar\,\left(\phi_1^2(x,\alpha_t)-\phi_2^2(x,\alpha_t)\right)\,\pd_x f(x,t)=2 V_s(x,t)\, \phi_1(x,\alpha_t)\phi_2(x,\alpha_t)\\
&\quad  -i V_p(x,t)\, \left(\phi_1^2(x,\alpha_t)+\phi_2^2(x,\alpha_t)\right)-i\hbar \left(\phi_2(x,\alpha_t)\,\pd_t\phi_1(x,\alpha_t)-\phi_1(x,\alpha_t)\,\pd_t\phi_2(x,\alpha_t)\right)
\end{split}
\end{equation}
Finally, noting that $V_s(x,t),\, V_p(x,t)\in\mbb{R}$ and $f(x,t)\in\mbb{R}$ Eqs.~\eref{eq18}-\eref{eq20} allow to fully determine the auxilliary potential matrix $\mbb{V}(x,t)$ and the phase $f(x,t)$. 

Separating real and imaginary parts in Eq.~\eref{eq20} is a simple exercise, however the resulting expressions are rather lengthy. Therefore, we will continue in the next section with two illustrative and analytically solvable examples.

\section{Illustrative examples \label{sec:ex}}

\subsection{Spatially homogeneous electric field \label{sec:ex1}}

We start with the simplest possible example. Imagine charged particles driven by a time-dependent, but spatially homogeneous vector field, and we simply have
\begin{equation}
\label{eq21}
A(x,\alpha_t) =\alpha_t\,.
\end{equation}
This situation has been recently analyzed in detail in the context of pair production \cite{Fillion2012} and the quantum work distribution \cite{Deffner2015}.

\subsubsection{Schr\"odinger dynamics}

To build intuition and as a point of reference we treat the corresponding Schr\"odinger case first. The ``unperturbed``, but driven dynamics \eref{eq06} becomes,
\begin{equation}
\label{eq22}
i \hbar\, \pd_t \psi_0(x,t)=\frac{1}{2 m}\left(-i \hbar\, \pd_x +\frac{\alpha_t}{c}\right)^2\,\psi_0(x,t)\,.
\end{equation} 
It is then easy to see that the instantaneous eigenstates are given by \cite{Deffner2015}
\begin{equation}
\label{eq23}
\phi(x,\alpha_t)=\ex{\frac{i}{\hbar} \left(\hbar\kappa+\frac{\alpha_t}{c}\right) \,x}\,,
\end{equation}
with the corresponding eigenenergies $\epsilon_\kappa(\alpha)=\left(\hbar\kappa+\alpha/c\right)^2/2m$, and $\kappa$ denotes the quantum number. Comparing Eq.~\eref{eq23} with Eq.~\eref{eq10} we identify $\beta(x,\alpha_t)=1$ and $\gamma(x,\alpha_t)=\left(c\,\hbar\kappa+\alpha_t\right) \,x/\hbar c$. Then, the determining equation for the phase $f(x,t)$ becomes,
$\pd^2_x f(x,t)=0$, for which a solution is given by
\begin{equation}
\label{eq25}
f(x,t)=a(t)+b(t) \,x\,.
\end{equation}
Substituting the instantaneous eigenstate \eref{eq23} and Eq.~\eref{eq25} into Eq.~\eref{eq11} we obtain
\begin{equation}
\label{eq26}
V(x,t)=-\hbar\, \left(\dot{a}(t)+\dot{b}(t)\, x\right)-\frac{\dot{\alpha}_t}{c}\,x+\frac{\hbar\,\alpha_t\,b(t)}{2 m c}+ \frac{\hbar^2\,b(t)^2}{2 m}+\frac{\hbar\,b(t)}{m}\,\left(\hbar \kappa+\frac{\alpha_t}{c}\right)\,,
\end{equation}
where the dot is again a short notation for a derivative with respect to time. Choosing
\begin{equation}
\label{eq27}
a(t)=0\quad \mathrm{and} \quad b(t)=0
\end{equation}
the phase and auxiliary potential simplify to read,
\begin{equation}
\label{eq28}
f(x,t)=0\quad\mathrm{and}\quad V(x,t)= -\frac{\dot{\alpha}_t}{c}\,x\,.
\end{equation}
From Eq.~\eref{eq28} it is obvious that for $\dot{\alpha}_0=0$ and $\dot{\alpha}_\tau=0$ the auxiliary potential $V(x,t)$ realizes the desired shortcut. It is interesting to note that the phase $f(x,t)$ vanishes. This, however, is to be expected as the solution of the ``unperturbed'' problem \eref{eq22} is fully determined by a time-dependent phase \cite{Deffner2015}.

\subsubsection{Dirac dynamics}

The solution of the analogous, time-dependent Dirac problem is mathematically more involved \cite{Deffner2015}. In general, a solution to the ``unperturbed'' equation,
\begin{equation}
\label{eq29}
i\hbar\, \pd_t\,\Psi_0(x,t)=\left[\left(-i\hbar c\,\pd_x+\alpha_t\right)\,\sigma_x+m c^2 \sigma_z\right]\,\Psi_0(x,t)\,,
\end{equation}
can only be written in terms of special functions \cite{Fillion2012,Deffner2015}. Nevertheless, we will see shortly that finding a shorcut to adiabaticity is a straight-forward exercise.

The instantaneous eigenspinor reads \cite{Deffner2015}
\begin{equation}
\label{eq30}
\Phi(x,\alpha_t)=\frac{\ex{i/\hbar\, \left(\hbar\kappa+\alpha_t/c\right) \,x}}{\sqrt{1 + \left(\sqrt{\Pi_t^2 + 1} - \Pi_t\right)^2 }}\,\begin{pmatrix}
1\\
 \sqrt{\Pi_t^2 + 1} - \Pi_t
\end{pmatrix},
\end{equation}
with $\Pi_t=(c\, \hbar\kappa +\alpha_t)/mc^2$ and the eigenenergies are $\epsilon_\kappa(\alpha)=\pm\sqrt{\left(c\, \hbar\kappa+\alpha_t\right)^2+(mc^2)^2}$. Therefore, Eq.~\eref{eq20} becomes
\begin{equation}
\label{eq31}
\frac{2}{\sqrt{1+\Pi^2_t}}\,\left(c \hbar\, \Pi_t\,\pd_x f(x,t)- V_s(x,t)\right)+i \left(2\, V_p(x,t)+\frac{\hbar\, \dot{\Pi}_t}{1+\Pi_t^2}\right)=0
\end{equation}
which is already separated into real and imaginary parts. Hence, the pseudoscalar potential is determined to read
\begin{equation}
\label{eq32}
V_p(x,t)=-\frac{\hbar\,\dot{\Pi}_t}{2+2\, \Pi_t^2}
\end{equation}
which vanishes for all parametrizations with $\dot{\alpha}_0=0$ and $\dot{\alpha}_\tau=0$, whereas the scalar potential is given by
\begin{equation}
\label{eq32a}
V_s(x,t)=c \hbar\, \Pi_t\,\pd_x f(x,t)\,.
\end{equation}
Substituting Eqs.~\eref{eq30}, \eref{eq32}, and \eref{eq32a} into Eq.~\eref{eq18} we obtain
\begin{equation}
\label{eq33}
V_t(x,t)=-m c\, \dot{\Pi}_t\,x-\hbar\, \pd_t f(x,t)-\hbar c\, \sqrt{1+\Pi_t^2}\,\pd_x f(x,t)\,.
\end{equation}
Now, choosing in complete analogy to the Schr\"odinger case the space-dependent phase to vanish, $f(x,t)\equiv 0$, we have $f(x,0)=0$ and $f(x,\tau)=0$, and $V_s(x,t)\equiv 0$. In conclusion, a shortcut to adiabaticity for the driven Dirac equation \eref{eq29} is facilitated by the auxiliary potential matrix,
\begin{equation}
\label{eq34}
\mbb{V}(x,t)= -m c\, \dot{\Pi}_t\,x\,\id_2-\frac{\hbar\,\dot{\Pi}_t}{2+2\, \Pi_t^2}\, \sigma_y=-\frac{\dot{\alpha}_t}{c}\,x\,\id_2-\frac{\hbar\,\dot{\alpha}_t}{2\left(\alpha_t+c\,\hbar\kappa+mc^2\right)}\, \sigma_y
\end{equation}
for all parametrizations with  $\dot{\alpha}_0=0$ and $\dot{\alpha}_\tau=0$. It is interesting to note that the $V_t(x,t)$-component is identical the auxiliary potential of the Schr\"odinger case \eref{eq28}, and the pseudoscalar potential $V_p(x,t)$ ensures the suppression of pair production at the final state. 

\subparagraph{Numerical verification and illustration}

Our analytical results \eref{eq34} can be easily verified. To this end, we solved  the ``unperturbed'', but driven Dirac equation \eref{eq29} as well as the fast-forwarded dynamics \eref{eq16} with the auxiliary potential \eref{eq34} numerically. The vector field \eqref{eq21} is parametrized by the protocol,
\begin{equation}
\label{eq34a}
\alpha_t=\si{\pi t/2 \tau}^2+1
\end{equation}
which fulfills $\dot{\alpha}_0=0$ and $\dot{\alpha}_\tau=1$. We further introduce the notation for the final state of the ``unperturbed'' dynamics \eref{eq29}, $\Psi_0(x,\tau)=(\psi_0^1,\psi_0^2)$, for the fast-forwarded state \eqref{eq16}, $\Psi(x,\tau)=(\psi^1_\mrm{FF},\psi^2_\mrm{FF})$, and for the instantaneous eigenstate, $\Phi(x,\alpha_\tau)=(\phi^1_\tau,\phi^2_\tau)$.

\begin{figure}
\centering
\includegraphics[width=.49\textwidth]{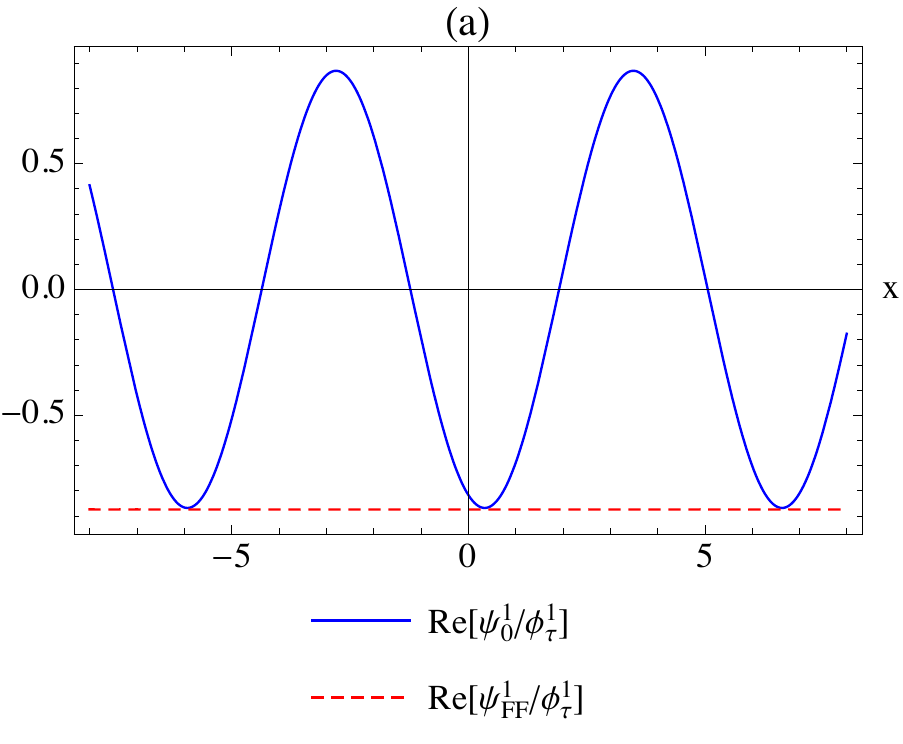} \hfill \includegraphics[width=.49\textwidth]{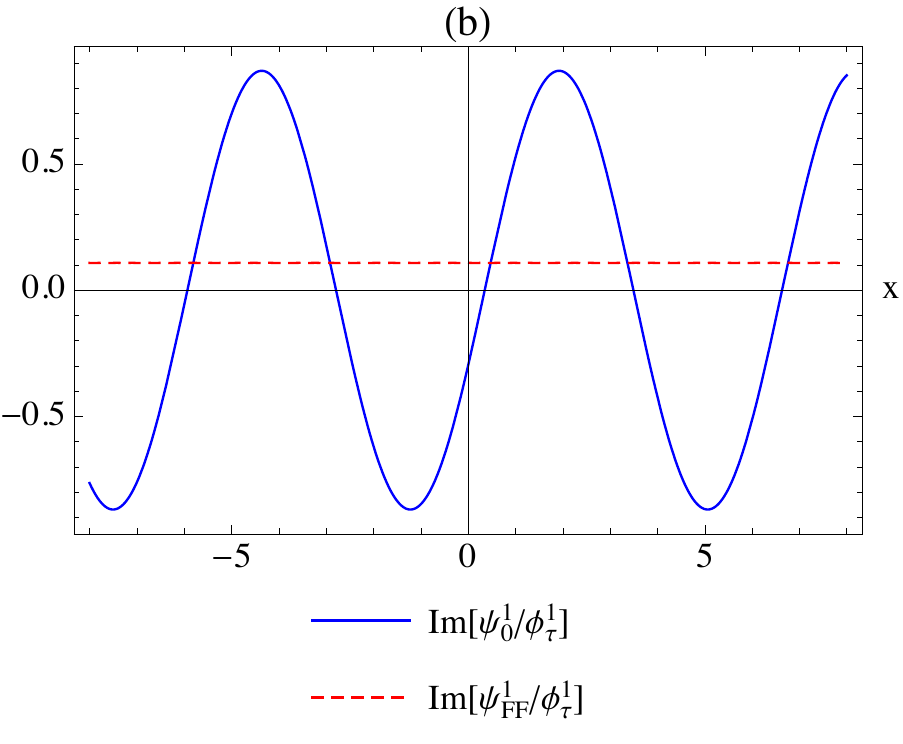}\\
\includegraphics[width=.49\textwidth]{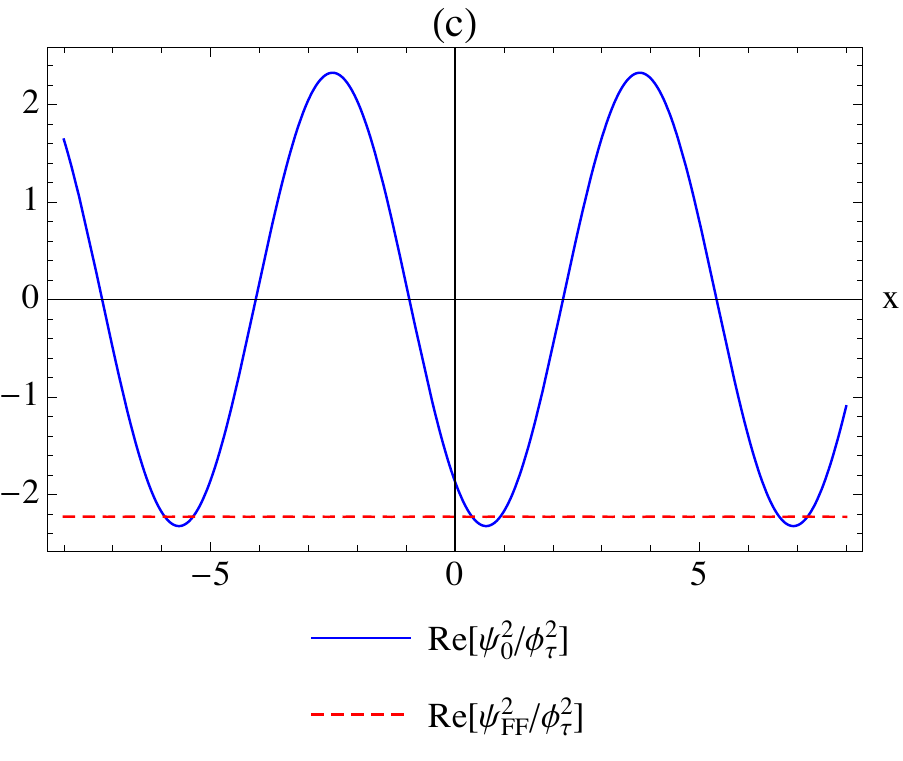} \hfill \includegraphics[width=.49\textwidth]{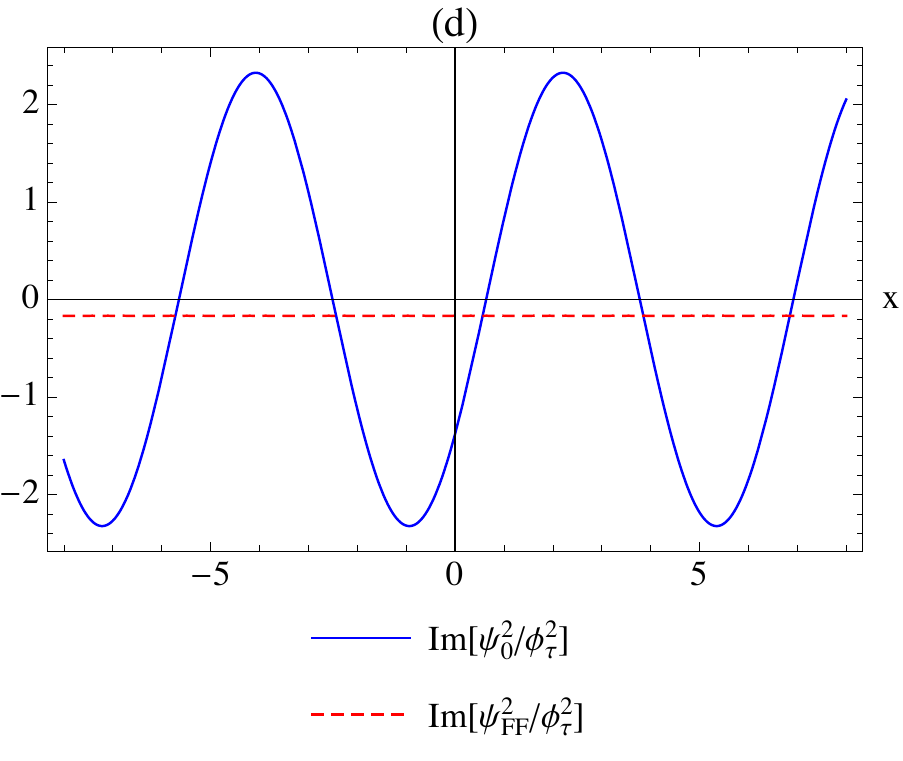}
\caption{\label{fig1} (color online) Illustration of the successfully fast-forwarded dynamics for spatially homogeneous vector field \eref{eq21} for the protocol \eref{eq34a}. Real (a) and imaginary (b) parts of the ratios of the first components, $\psi_0^1/\phi^1_\tau$ and $\psi_\mrm{FF}^1/\phi^1_\tau$, and real (c) and imaginary (d) parts of the second components, $\psi_0^2/\phi^2_\tau$ and $\psi_\mrm{FF}^2/\phi^2_\tau$. Parameters are $m=1$, $c=1$, $\hbar=1$, $\kappa=0$, and $\tau=1$.}
\end{figure}

In Fig.~\ref{fig1} we plot the real and imaginary parts of the ratios $\psi_0^1/\phi^1_\tau$, $\psi_0^2/\phi^2_\tau$, $\psi_\mrm{FF}^1/\phi^1_\tau$, and $\psi_\mrm{FF}^2/\phi^2_\tau$. We observe that these ratios are independent of position $x$ for the fast-forwarded state, whereas the solution of the ``unperturbed'' dynamics \eref{eq29} yields a strong dependence on $x$. This is in full agreement with the analytical treatment, for which we expect the fast-forwarded state $\Psi(x,\tau)$ to be identical to an eigenspinor up to a space-independent phase. %\footnote{Note that generally and in complete analogy to Schr\"odinger dynamics the time-dependent Dirac equation exhibits dynamical and geometric phase \cite{Anandan1993,Arunagiri2011}. For the present $(1+1)$-dimensional Dirac equation the geometric part is given by a space-independent, unitary $(2\times 2)$-matrix, whose components can be determined from the constant real and imaginary parts of the ratios $\psi_\mrm{FF}^1/\phi^1_\tau$, and $\psi_\mrm{FF}^1/\phi^2_\tau$.}

\subsection{Linear electric field \label{sec:ex2}}

As a second example we consider charged particles driven by a linear electric field. The electromagnetic vector field is then given by,
\begin{equation}
\label{eq35}
A(x,\alpha_t) =\alpha_t\,x\,.
\end{equation}
As before, we will first solve the Schr\"odinger case, and then analyze the fast-forwarded Dirac dynamics. 

\subsubsection{Schr\"odinger dynamics}

The ``unperturbed'', but driven Schr\"odinger equation \eref{eq06} becomes for Eq.~\eref{eq35},
\begin{equation}
\label{eq36}
 i \hbar\, \pd_t \psi_0(x,t)=\frac{1}{2 m}\left(-i \hbar\, \pd_x +\frac{\alpha_t}{c}\, x\right)^2\,\psi_0(x,t)\,.
\end{equation}
It is easy to see that an instantaneous solution of Eq.~\eref{eq36} can be written as,
\begin{equation}
\label{eq37}
\phi(x,\alpha_t)=\ex{-\frac{i}{\hbar}\left(\frac{\alpha_t}{2 c}\, x^2 - \hbar \kappa\,x\right)}
\end{equation}
with the instantaneous eigenenergies $\epsilon_\kappa=(\hbar\kappa)^2/2m$. Accordingly, we identify $\beta(x,\alpha_t)=1$ and $\gamma(x,\alpha_t)=-\alpha_t x^2/2\hbar c+\kappa x$. Therefore, the differential equation for the phase $f(x,t)$ again simplifies to, $\pd_x^2 f(x,t)=0$, for which we write the solution as
\begin{equation}
\label{eq38}
f(x,t)=a(t)+b(t)\, x\,.
\end{equation}
Substituting Eqs.~\eref{eq36} and \eref{eq28} into Eq.~\eref{eq11} we have,
\begin{equation}
\label{eq39}
V(x,t)=\frac{\dot{\alpha}_t}{2 c}\,x^2-\left(\frac{\hbar}{2 mc}\,b(t)\,\alpha_t+\hbar\,\dot{b}(t)\right)\,x+\frac{\hbar^2\kappa}{m}\,b(t)+\frac{\hbar^2}{2m}\,b(t)^2-\hbar\,\dot{a}(t)\,.
\end{equation}
In order to have a shortcut, we have to demand that phase, $f(x,t)$ and auxiliary potential $V(x,t)$, vanish for $t=0$ and $t=\tau$. This requirement is met for the particular choice $a(t)=0$ and $b(t)=0$, and we have
\begin{equation}
\label{eq40}
V(x,t)=\frac{\dot{\alpha}_t}{2 c}\,x^2
\end{equation}
which facilitates a shortcut to adiabaticity for all protocols with $\dot{\alpha}_0=0$ and $\dot{\alpha}_\tau=0$.

\subsubsection{Dirac dynamics}

As a final example we now consider the corresponding Dirac problem. The driven Dirac equation \eref{eq13} becomes 
\begin{equation}
\label{eq41}
i\hbar\, \pd_t\,\Psi_0(x,t)=\left[\left(-i\hbar c\,\pd_x+\alpha_t\,x\right)\,\sigma_x+m c^2 \sigma_z\right]\,\Psi_0(x,t)\,,
\end{equation}
for which the instantaneous eigenstates are given by
\begin{equation}
\label{eq42}
\Phi(x,\alpha_t)=\frac{\ex{-i/\hbar\,\left(\alpha_t x^2/2 c - \hbar \kappa\,x\right)}}{\sqrt{1 + \left(\sqrt{\Lambda^2 + 1} - \Lambda\right)^2 }}\,\begin{pmatrix}
1\\
 \sqrt{\Lambda^2 + 1} - \Lambda
\end{pmatrix},
\end{equation}
where $\Lambda=\hbar\kappa/mc$. In \ref{sec:app} we summarize how to determine $\Phi(x,\alpha_t)$, and for the three dimensional case the derivation can be found in Ref.~\cite{Moshinsky1989}. Accordingly, Eq.~\eref{eq20} simplifies to
\begin{equation}
\label{eq43}
\frac{1}{\sqrt{\Lambda^2+1}}\,\left(\hbar c \Lambda\,\pd_x f(x,t)-V_s(x,t)\right)+i V_p(x,t)=0\,,
\end{equation}
from which we determine $V_p(x,t)\equiv 0$ and $V_s(x,t)=\hbar c \Lambda\,\pd_x f(x,t) $. Thus, $V_t(x,t)$ can be written as
\begin{equation}
\label{eq44}
V_t(x,t)=\frac{\dot{\alpha}_t}{2 c}\,x^2-\hbar\, \pd_t f(x,t)-\hbar c\sqrt{1+\Lambda^2} \,\pd_x f(x,t)\,.
\end{equation}
Again choosing $f(x,t)\equiv 0$ a shortcut to adiabaticity is induced by the auxiliary potential matrix
\begin{equation}
\label{eq45}
\mbb{V}(x,t)=\frac{\dot{\alpha}_t}{2 c}\,x^2 \,\id_2
\end{equation}
for all protocols with $\dot{\alpha}_0=0$ and $\dot{\alpha}_\tau=0$. Note that in the case of a linear electric field \eref{eq35} the auxiliary potentials for Schr\"odinger dynamics \eref{eq40} and for Dirac dynamics \eref{eq44} are identical and no peseudscalar potential is necessary to suppress pair creation at the final state. 

\subparagraph{Numerical verification and illustration}

\begin{figure}
\centering
\includegraphics[width=.49\textwidth]{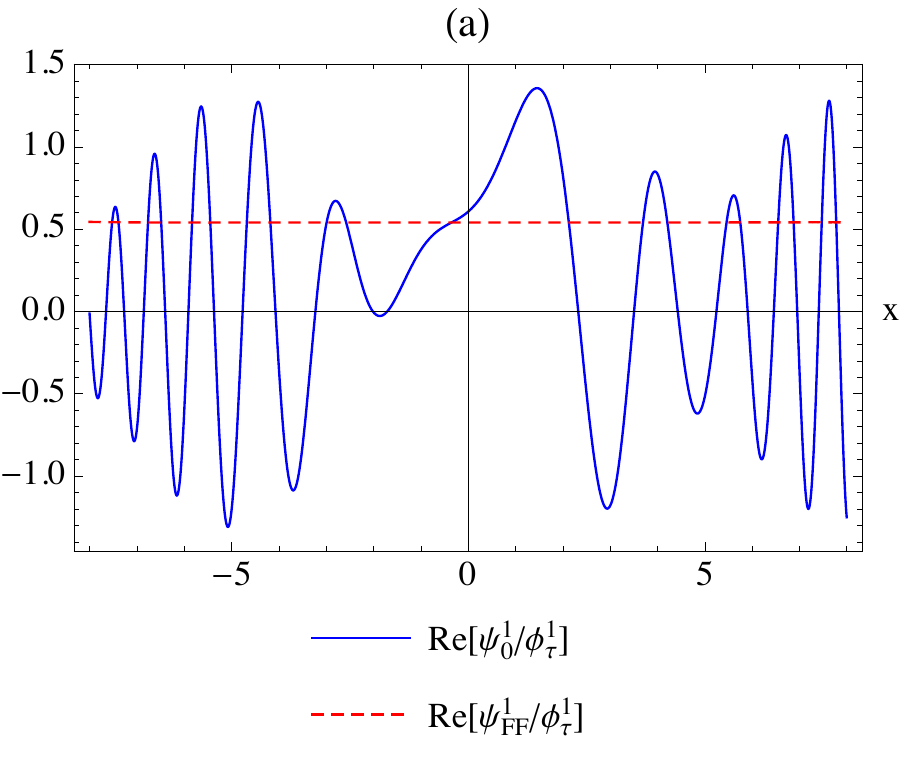} \hfill \includegraphics[width=.49\textwidth]{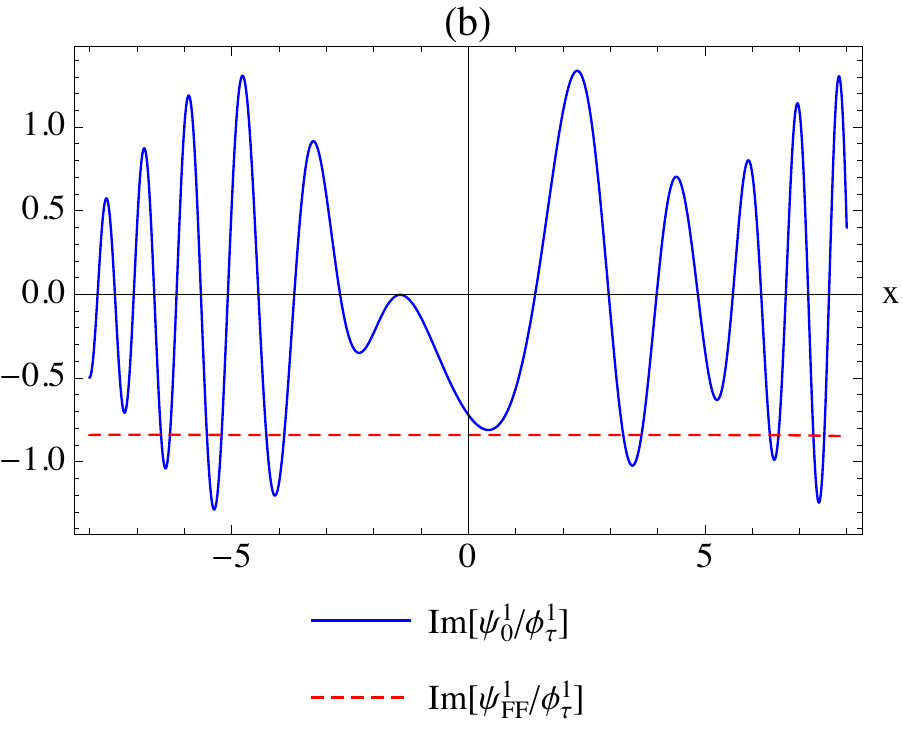}\\
\includegraphics[width=.49\textwidth]{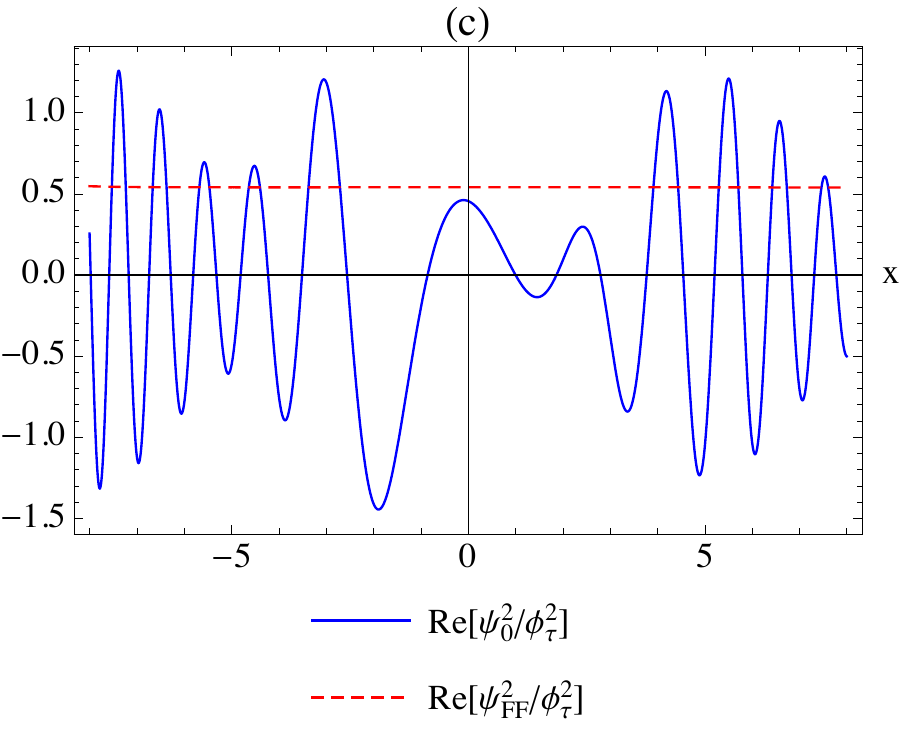} \hfill \includegraphics[width=.49\textwidth]{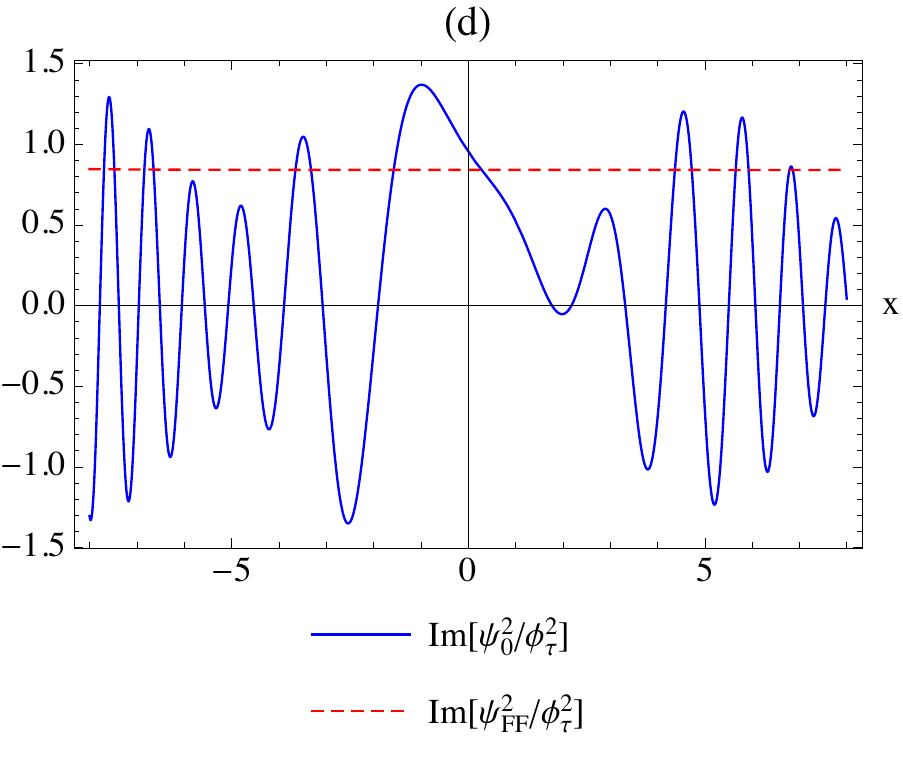}
\caption{\label{fig2} (color online) Illustration of the successfully fast-forwarded dynamics for linear vector field \eref{eq35} for the protocol \eref{eq34a}. Real (a) and imaginary (b) parts of the ratios of the first components, $\psi_0^1/\phi^1_\tau$ and $\psi_\mrm{FF}^1/\phi^1_\tau$, and real (c) and imaginary  (d) parts of the second components, $\psi_0^2/\phi^2_\tau$ and $\psi_\mrm{FF}^2/\phi^2_\tau$. Parameters are $m=1$, $c=1$, $\hbar=1$, $\kappa=0$, and $\tau=1$.}
\end{figure}

In Fig.~\ref{fig2} we plot again real and imaginary parts  of the ratios $\psi_0^1/\phi^1_\tau$, $\psi_0^2/\phi^2_\tau$, $\psi_\mrm{FF}^1/\phi^1_\tau$, and $\psi_\mrm{FF}^2/\phi^2_\tau$ for the sinusoidal protocol \eref{eq34a}. As before in Fig.~\ref{fig1} we observe that the fast-forwarded dynamics results only in a space-independent phase, whereas the final state for ``unperturbed'' equation \eref{eq41} lies far from the instantaneous eigenspinor. 

\section{Concluding remarks \label{sec:con}}

In the present work we have generalized the fast-forward technique to shortcuts to adiabaticity for Dirac dynamics. For the Dirac equation such a shortcut not only suppresses parasitic nonequilibrium excitations, but also hinders the production of pairs. Therefore, our results could be applied in particle physics for problems, in which one is interested in loss-free  transport of charged particles, as well as in condensed matter physics, where one is interested in the dissipationless control of Dirac materials. As an illustration of the technique we have worked out two analytically solvable examples, namely charged particles in spatially homogeneous and linear fields. We have found that the scalar component of the auxiliary potential matrix is identical to the auxiliary potential in the analogous Schr\"odinger problem. However, we have also seen that in order to facilitate a shortcut in Dirac dynamics pseudoscalar potentials are important. 

As a final observation we remark that in all examples we were able to choose the fast-forward phase to vanish at all times. Hence, for the present examples the auxiliary potential matrix is identical to the counterdiabatic field \cite{Deffner2014}, and the fast-forwarded Dirac spinor does not stray from the adiabatic manifold of the ``unperturbed'' Hamiltonian. 

\ack{It is a pleasure to thank Avadh Saxena, who got me interested in Dirac dynamics, and Bart\l omiej Gardas for insightful discussions.  SD acknowledges financial support by the U.S. Department of Energy through a LANL Director's Funded Fellowship.}

\appendix

\section{Dirac particles in linear electric field \label{sec:app}}

The instantaneous eigenvalue problem for the driven Dirac equation \eref{eq41} reads separated into its components 
\begin{equation}
\label{eqa1}
\begin{split}
\left(\epsilon-m c^2\right)\phi_1(x,\alpha)&=-i\hbar c\,\pd_x\phi_2(x,\alpha)+\alpha x\,\phi_2(x,\alpha)\\
\left(\epsilon+m c^2\right)\phi_2(x,\alpha)&=-i\hbar c\,\pd_x\phi_1(x,\alpha)+\alpha x\,\phi_1(x,\alpha)\,.
\end{split}
\end{equation}
Multiplying the first equation by $(\epsilon+m c^2)$ and substituting the second one into the first we obtain
\begin{equation}
\label{eqa2}
\left(\epsilon^2-m^2 c^4\right)\phi_1(x,\alpha)=\left(-i\hbar c\,\pd_x+\alpha x\right)^2\,\phi_1(x,\alpha)\,,
\end{equation}
which is similar to the corresponding Schr\"odinger equation \eref{eq36}. Accordingly, a solution is given by
\begin{equation}
\label{eqa3}
\phi_1(x,\alpha)=\ex{-\frac{i}{\hbar}\left(\frac{\alpha_t}{2 c}\, x^2 - \hbar \kappa\,x\right)}
\end{equation}
for which the eigenenergies read
\begin{equation}
\label{eqa4}
\epsilon_\kappa(\alpha)=\pm \sqrt{\left(\hbar\kappa c\right)^2+\left(m c^2\right)^2}\,.
\end{equation}
Re-substituting Eq.~\eref{eqa4} into Eq.~\eref{eqa1} we obtain the second component as
\begin{equation}
\label{eqa5}
\phi_2(x,\alpha)=\frac{\hbar\kappa c}{\sqrt{\left(\hbar\kappa c\right)^2+\left(m c^2\right)^2}+m c^2}\, \ex{-\frac{i}{\hbar}\left(\frac{\alpha_t}{2 c}\, x^2 - \hbar \kappa\,x\right)}\,.
\end{equation}
Combining $\phi_1(x,\alpha)$ and $\phi_2(x,\alpha)$, and normalizing in momentum space we have the instantaneous eigenspinors \eref{eq42}, see also Ref.~\cite{Moshinsky1989,Deffner2015}

\bibliographystyle{unsrt}
\bibliography{FF_Dirac_revised}

\end{document}